\documentclass[a4paper]{PoS}

\title{Hadronic $B$ decay reconstruction in early Belle II data}

\ShortTitle{Hadronic $B$ decay reconstruction in early Belle II data}

\author{\speaker{Eldar Ganiev}\footnote{On behalf of the Belle II collaboration.}\\
        University of Trieste and INFN, Trieste, Italy\\
        E-mail: \email{eldar.ganiev@ts.infn.it}}

\author{Niharika Rout\\
        Indian Insitute of Technology Madras, Chennai, India\\
        E-mail: \email{niharikarout@physics.iitm.ac.in}}

\author{Benedikt Wach\\
		Max Planck Insitute for Physics, Munich, Germany\\
		E-mail: \email{wach@mpp.mpg.de}}

\abstract{Belle II is an experiment designed to study billions of $\tau$-lepton, $b$- and $c$-quark decays observed with low background in asymmetric-energy electron-positron collisions at the SuperKEKB $B$-factory. In March 2019, the newly completed Belle II started operating and collected its first physics data reaching 10 fb$^{-1}$ to date. We report the reconstruction of prominent signals from various hadronic $B$ decays including $B^{-}\to D^{(*)0}\pi^{-}$, $B^{0}\to D^{-}K^{+}$, and $B^{0}\to K^{+}\pi^{-}$ in the first data set corresponding to 5.15 fb$^{-1}$. These results show a remarkable level of early understanding of detector performance.}

\FullConference{18th International Conference on B-Physics at Frontier Machines - Beauty2019 -\\
		29 September / 4 October, 2019\\
		Ljubljana, Slovenia}

\begin{document}

\section{Introduction}

The Standard Model (SM) describes accurately thousands of measurements up to the TeV-energies explored so far. However, several open questions suggest that the SM should be completed by a more general theory that extends it to higher energies. Determining the theory that completes the SM is the main goal of today's high-energy physics. Flavor physics offers powerful approach in indirect searches for SM extensions. 

The Belle II hadronic $B$ decay program is expected to have a key role with precise measurements of the $\alpha/\phi_{2}$ and $\gamma/\phi_{3}$ angles of the Cabibbo-Kobayashi-Maskawa (CKM) triangle, tests of non-SM charge-parity violation (CPV) in penguin $b\to d$ and $b\to s$ transitions in $B^{0}\to\eta 'K^{0}$ and $B^{0}\to\phi K^{0}$ decays [1], and many other processes. 

The Belle II experiment at the SuperKEKB collider is designed to reconstruct decays of billions of heavy-flavor particles and $\tau$-leptons. SuperKEKB is a $B$-factory that operates asymmetric-energy beams of 7 and 4 GeV for $e^{-}$ and $e^{+}$, respectively. The design luminosity is \linebreak 8 $\times$ 10$^{35}$ cm$^{-2}$s$^{-1}$, which should be achieved through a novel low-emittance nanobeam scheme [2]. The Belle II detector is a large-solid-angle magnetic spectrometer including a vertex detector, a central drift chamber, two dedicated particle-identification (PID) systems, an electromagnetic calorimeter, and outer detectors for $K^{0}_{L}$ and muon detection. Construction was completed in late 2018 and data were taken throughout 2019, reaching a peak luminosity of 1.1 $\times$ 10$^{34}$ cm$^{-2}$s$^{-1}$ and integrating 9.2 fb$^{-1}$ at the $\Upsilon$(4S) and 0.8 fb$^{-1}$ off resonance by December 2019. Results shown here are restricted to the 5.15 fb$^{-1}$ of $\Upsilon$(4S) data taken from March to July. We reconstruct well-known $B$ decays by using basic tools and implementing baseline selections to gain insight on the performance of our novel detector. We first identify the subset of observables that show consistency with simulation. Then, the reconstruction strategy is fully developed on simulated data and is finally applied to experimental data.
 
\section{$B$ meson reconstruction and background suppression}

Electron and positron collide to produce the $\Upsilon$(4S) resonance, which decays almost exclusively to a $B\bar{B}$ pair (signal), and pairs of lighter quarks (continuum background) with a four times higher rate. The $B$ meson has a distinctive invariant mass peak over the smooth distribution of continuum events. In addition to that, the well-known collision energy offers two efficient discriminating variables, beam-constrained mass and energy difference, defined as
\begin{equation}
 M_{bc} \equiv \sqrt{E^{2}_{beam}/c^4 - |p_{B}/c|^{2}}, \;\;\;\;\; \Delta E \equiv E_{beam} - E_{B},
\end{equation} 
where $E_{B}$ and $p_{B}$ are the reconstructed energy and momentum of $B$ meson candidates in the center-of-mass (CM) frame, and $E_{beam}$ is the beam energy in the CM frame. Signal events tend to cluster in the (-0.05, 0.05) GeV region of $\Delta E$ and the (5.27, 5.29) GeV/$c^{2}$ region of $M_{bc}$.

We reconstruct the following decay modes (charge-conjugated modes are implied):
\begin{itemize}
\item $ B^{-} \to D^{0}(\to K^{-}\pi^{+},K^{-}\pi^{+}\pi^{0},K^{-}\pi^{+}\pi^{-}\pi^{+}) \pi^{-}$;
\item $ B^{-} \to D^{0}(\to K^{-}\pi^{+},K^{-}\pi^{+}\pi^{0},K^{-}\pi^{+}\pi^{-}\pi^{+}) \rho^{-}(\to \pi^-\pi^0) $;
\item $ B^{-} \to D^{*0}[\to D^{0}(\to K^{-}\pi^{+},K^{-}\pi^{+}\pi^{0},K^{-}\pi^{+}\pi^{-}\pi^{+})\pi^{0}]\pi^{-} $;
\item $ \overline{B}^{0} \to D^{*+}[\to D^{0}(\to K^{-}\pi^{+},K^{-}\pi^{+}\pi^{0},K^{-}\pi^{+}\pi^{-}\pi^{+})\pi^{+}]\pi^{-} $;
\item $ \overline{B}^{0} \to D^{+}[\to K^{-}\pi^{+}\pi^{+}, K_{S}^{0}(\to \pi^+\pi^-)\pi^{+}] \pi^{-}$;
\item $ \overline{B}^{0} \to D^{+}[\to K^{-}\pi^{+}\pi^{+}, K_{S}^{0}(\to \pi^+\pi^-)\pi^{+}] \rho^{-}(\to \pi^{-}\pi^0) $;
\item $ B^{-} \to D^{0}(\to K^{-}\pi^{+},K^{-}\pi^{+}\pi^{0},K^{-}\pi^{+}\pi^{-}\pi^{+}) K^{-}$;
\item $ \overline{B}^{0} \to D^{+}[\to K^{-}\pi^{+}\pi^{+}, K_{S}^{0}(\to \pi^+\pi^-)\pi^{+}] K^{-}$;
\item $ \overline{B}^{0} \to K^{-}\pi^{+}$.
\end{itemize}
In the following, we assume that $B$ identifies both $B^{+}$ and $B^{0}$, $D^{(*)}$ identifies ${D^{(*)}}^{+}$ and ${D^{(*)}}^{0}$, $\rho$ identifies $\rho^{+}$, $\pi$ identifies $\pi^{+}$ and $K$ identifies $K^{+}$.

The signal-to-background ratio at production varies from approximately 10$^{-3}$ (for $b\to c$ channels) to approximately 10$^{-6}$ (for charmless $b\to u, d, s$ channels). In addition, continuum events can mimic the final-state features of signal. Belle II adopts several techniques to suppress continuum based on the experience of the BaBar, Belle, CLEO and Argus experiments. Most of these techniques exploit the shape differences between continuum and $B\overline{B}$ events. In a continuum event, the lighter quarks are produced with momentum of about 5 GeV/$c$ and tend to fragment into two back-to-back jets of collimated light hadrons. $B$ meson pairs are produced almost at rest in the $\Upsilon$(4S) frame, as the $\Upsilon$(4S) mass exceeds only slightly the $B\overline{B}$-production threshold. Therefore, the $B$ decay products are distributed isotropically in the $\Upsilon$(4S) rest frame.

\begin{figure}[!htb]
\begin{center}
\includegraphics[width=8cm]{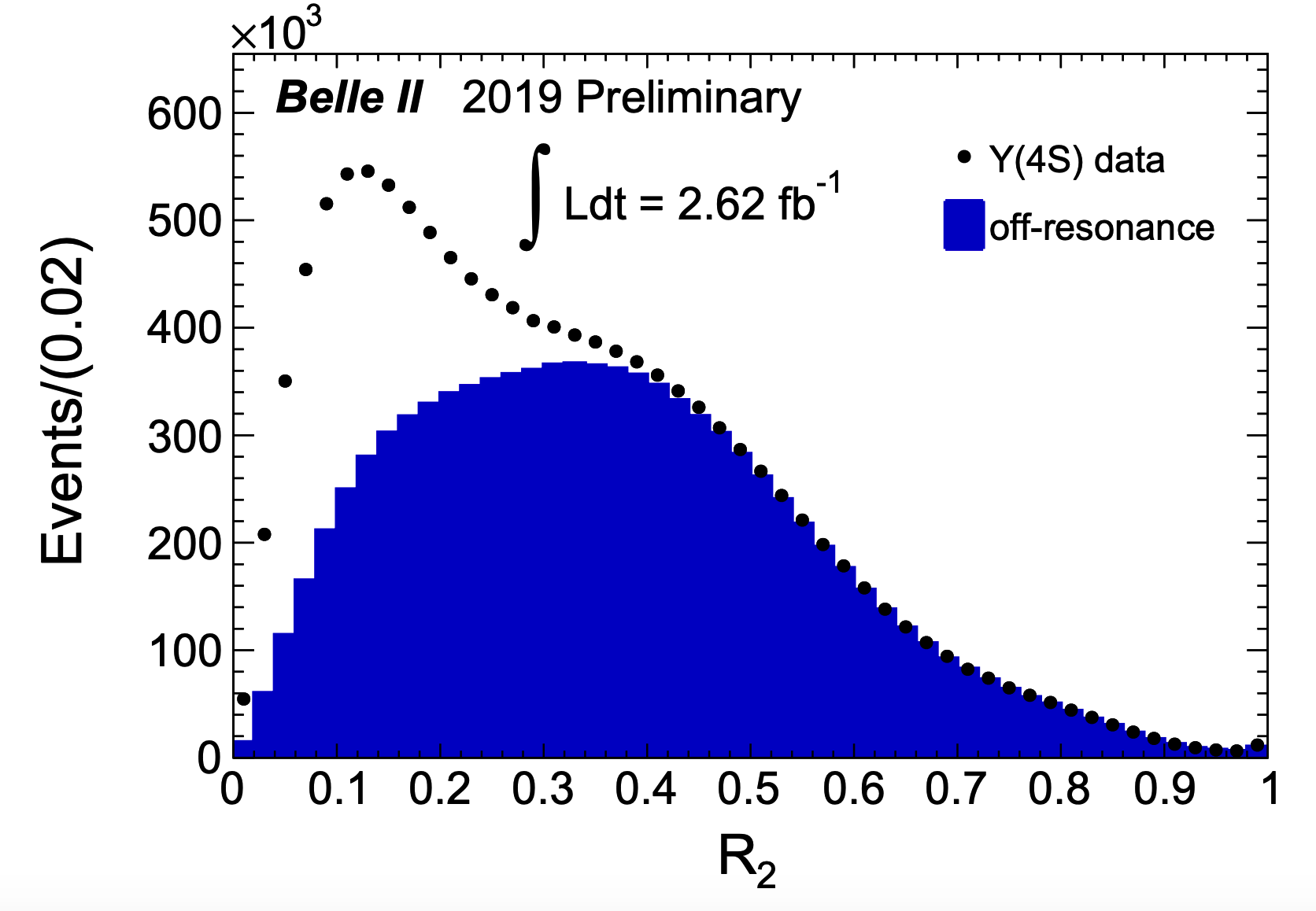}
\end{center}
 \caption{$R_{2}$ distributions for $\Upsilon$(4S) and off-resonance data.}
 \label{figr2}
\end{figure}

Fox-Wolfram moments [3] are typically used to exploit these features: given total $N$ particles in an event with momenta $\textbf{p}^{*}_{i}$ in the $\Upsilon$(4S) frame, the $l$th order Fox-Wolfram moment $H_{l}$ is defined as 
 \begin{equation}
  H_{l} = \sum_{i,j}^{N}\frac{|p^{*}_{i}|\cdot|p^{*}_{j}|}{s}\cdot P_{l}(cos\theta^{*}_{i,j}),
 \end{equation}  
where $\theta^{*}_{i,j}$ is the angle between $p^{*}_{i}$ and $p^{*}_{j}$, and $P_{l}$ is the $l$th order Legendre polynomial. The normalized ratio $R_{2}$ = $H_{2}/H_{0}$ is a simple quantity that already offers strong separation between signal and background (Fig.\ref{figr2}) and is used in $B\to D^{(*)}\pi$ and $B\to D^{(*)}\rho$ reconstruction. Because backgrounds outnumber the more suppressed of the signal decays by orders of magnitude, a more powerful discriminating method, fast boosted decision tree (FBDT), is used. FBDT combines nonlinearly 20+ kinematic, decay-time, PID, and topology variables to maximize the signal-to-background separation. FBDT classifier training is performed on ensembles of independent simulated samples.

\section{$B$ to charm decay results}

\begin{figure}[!htb]
\begin{center}
\begin{tabular}{c c}
\includegraphics[width=7cm]{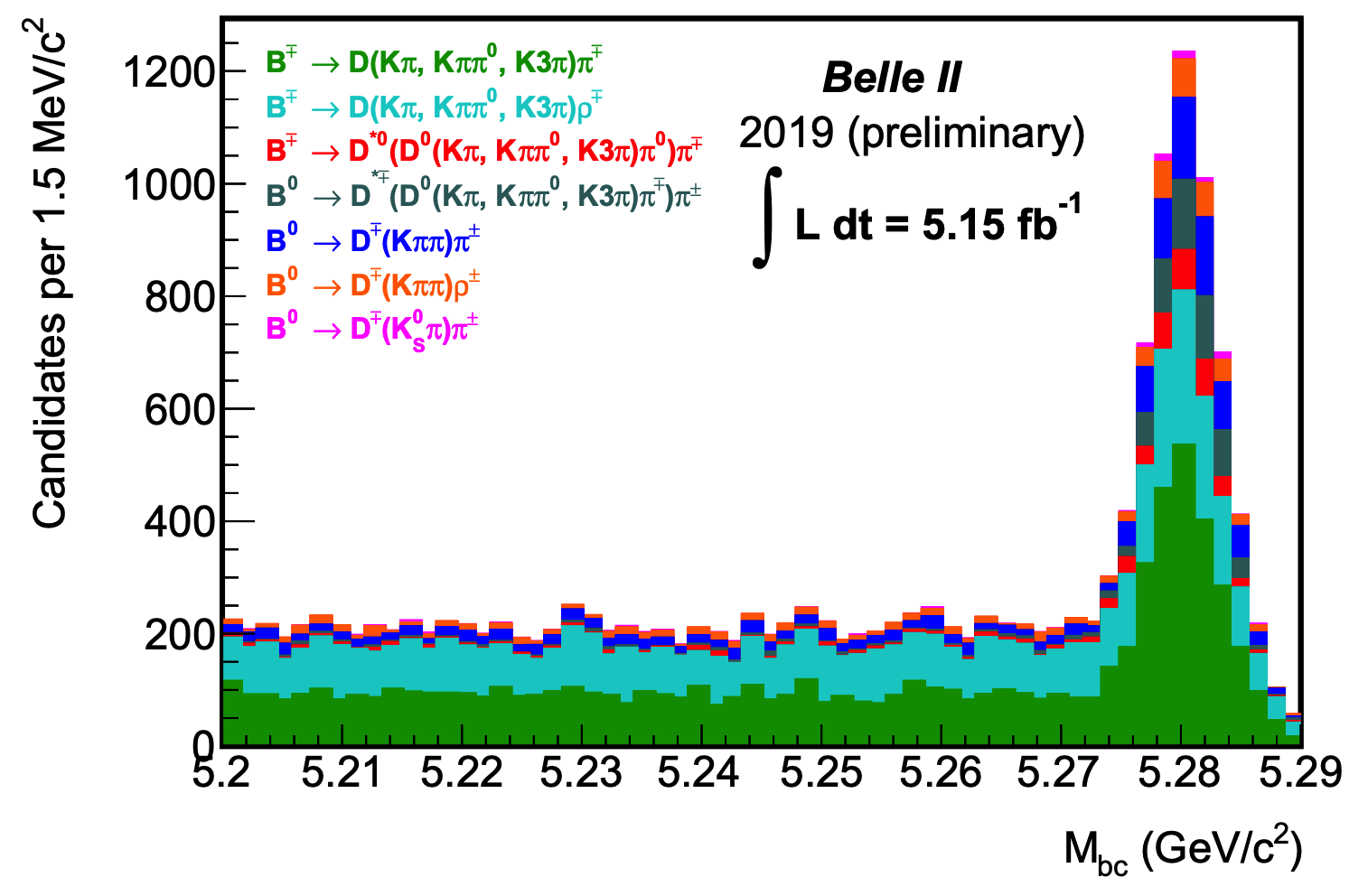} &
\includegraphics[width=7cm]{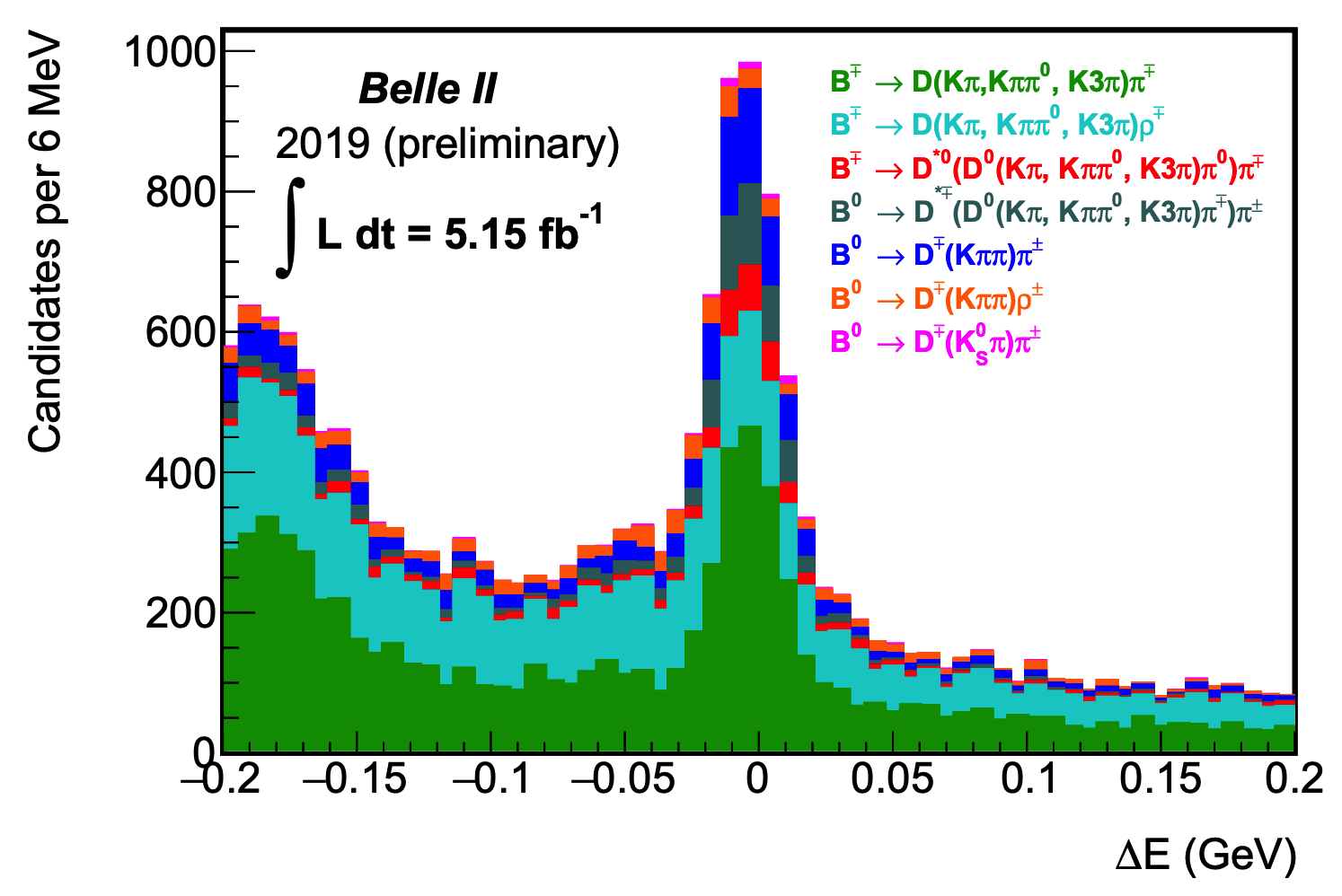} \\
\end{tabular}
\end{center}
 \caption{Distributions of $M_{bc}$ (left) and $\Delta$E (right) for $B\to D^{(*)}h$ ($h$ = $\pi$ or $K$) candidates reconstructed in 5.15 fb$^{-1}$ of collision data. The $M_{bc}$ distribution is obtained by restricting the data to the (-0.05, 0.05) GeV range of $\Delta E$; the $\Delta E$ distribution is obtained by restricting the data to the (5.27, 5.29) GeV/$c^{2}$ range of $M_{bc}$. A requirement on $R_{2}$ is applied to suppress the continuum background.}
 \label{stack}
\end{figure}

$B\to Dh$ modes offer useful properties to validate the detector performance, as they are abundant and involve charged and neutral final-state particles, multivertex topologies and long-lived final states. We first take the reconstruction primitives, e.g. tracks, and apply simple track-quality selection, mainly aimed to reject the beam background. These primitives are then combined into intermediate-resonance candidates ($D$, $\rho$), which are required to meet simple invariant-mass conditions. Finally, reconstructed particles are combined into $B$ mesons. The distributions of $M_{bc}$ and $\Delta E$ for various $B\to D^{(*)}\pi$ and $B\to D^{(*)}\rho$ decay modes are shown in Fig. \ref{stack}. Approximately 4500 events are observed in 5.15 fb$^{-1}$ [4]. 

\begin{figure}[!htb]
\begin{center}
\begin{tabular}{c c}
\includegraphics[width=7cm]{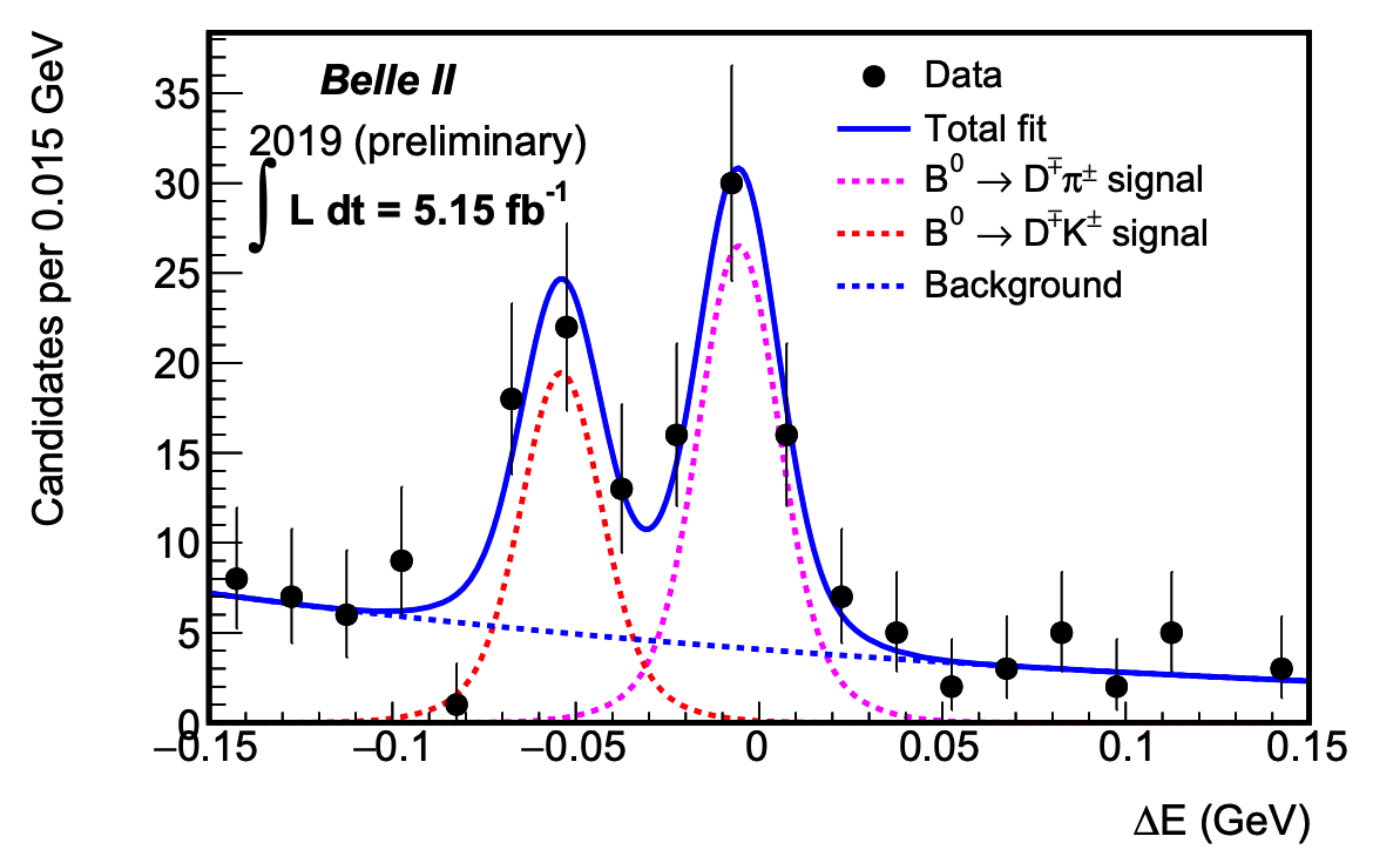} &
\includegraphics[width=6.5cm]{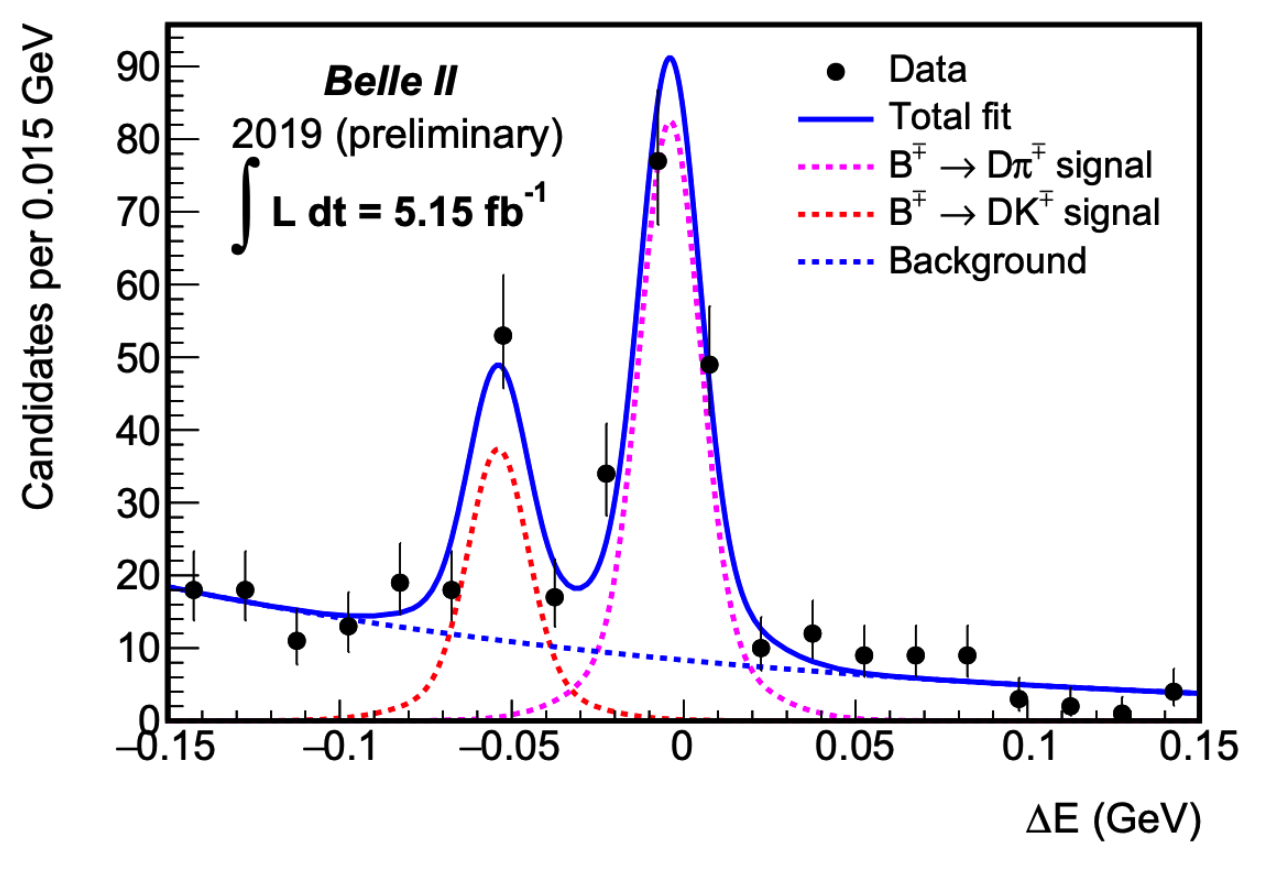} \\
\end{tabular}
\end{center}
 \caption{Distributions of $\Delta E$ for $B^{0}\to D^{-}h^{+}$ (left) and $B^{-}\to D^{0}h^{-}$ (right) ($h$ = $\pi$ or $K$) candidates reconstructed in 5.15 fb$^{-1}$ of collision data and restricted to the region $M_{bc}$ > 5.27 GeV/$c^{2}$. The projection of an unbinned maximum likelihood fit is overlaid. A FBDT requirement is applied to suppress continuum. In addition, a requirement on prompt-kaon PID enriches the sample with $B\to DK$ events.}
 \label{bdk}
\end{figure}

Furthermore, we search for the $B\to DK$ decay, which is the main channel used for the measurement of $\gamma/\phi_{3}$, and its observation offers important information for assessing PID and continuum-suppression performance. In addition to the above general requirements, we apply a requirement on the PID of the prompt kaon to enrich the sample with $B\to DK$ candidates. We obtain two adjacent peaking structures: the peak centered at $\Delta E$ $\approx$ 0 GeV is composed by the dominant $B\to D\pi$ decays reconstructed with a misidentified pion, the other peak at $\Delta E$ $\approx$ -0.5 GeV represents $B\to DK$ decays. The $\Delta$E distributions for $B^{0}\to D^{-}K^{+}$ and $B^{-}\to D^{0}K^{-}$ are shown in \linebreak Fig. \ref{bdk}. A total of 39 $\pm$ 8 $B^{0}\to D^{-}K^{+}$ and 53 $\pm$ 9 $B^{-}\to D^{0}K^{-}$ signal decays are observed in \linebreak 5.15 fb$^{-1}$ [4]. 

\section{First charmless $B$ decays}

\begin{figure}[!htb]
\begin{center}
\begin{tabular}{c c}
\includegraphics[width=7cm]{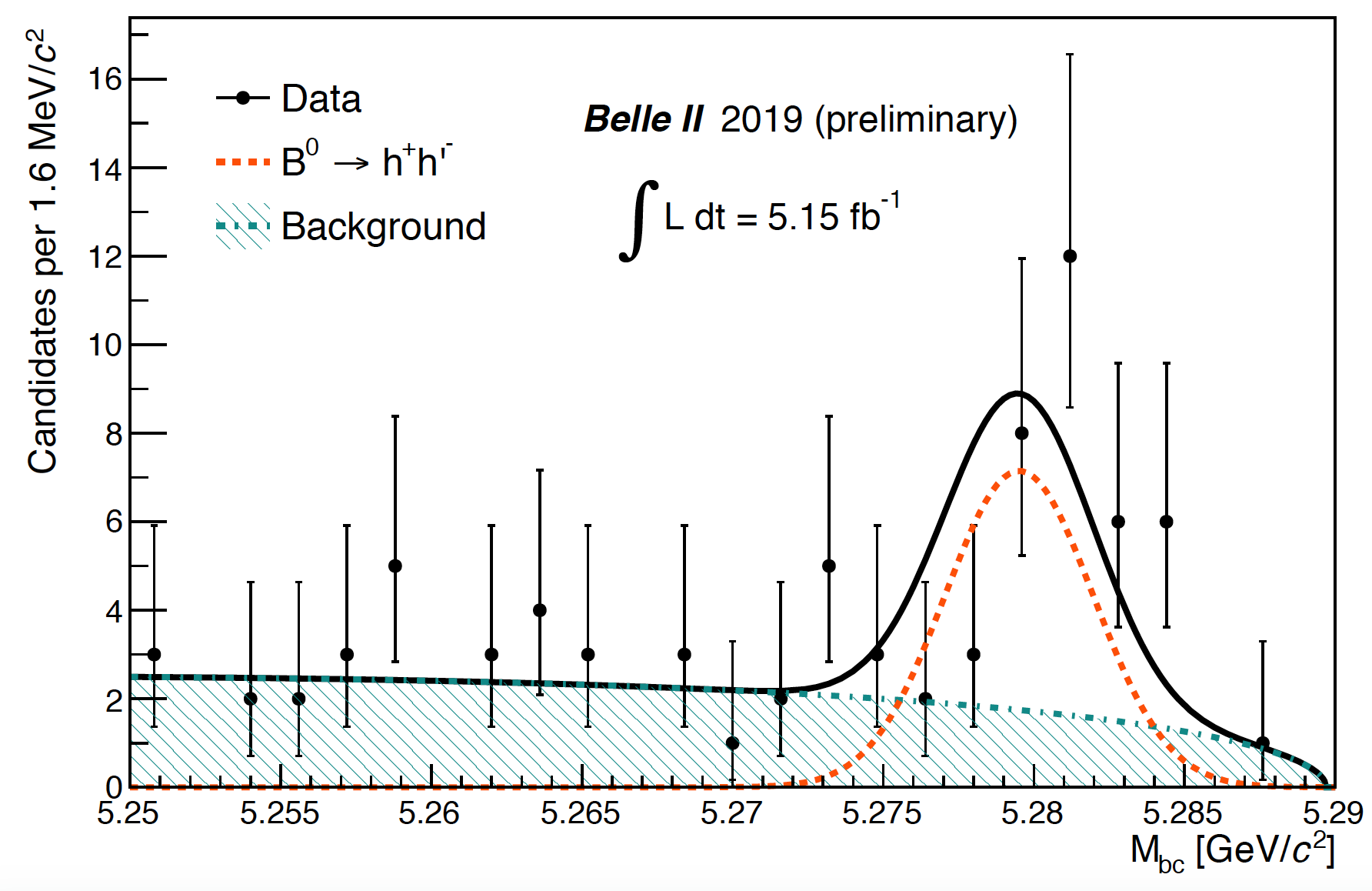} &
\includegraphics[width=7cm]{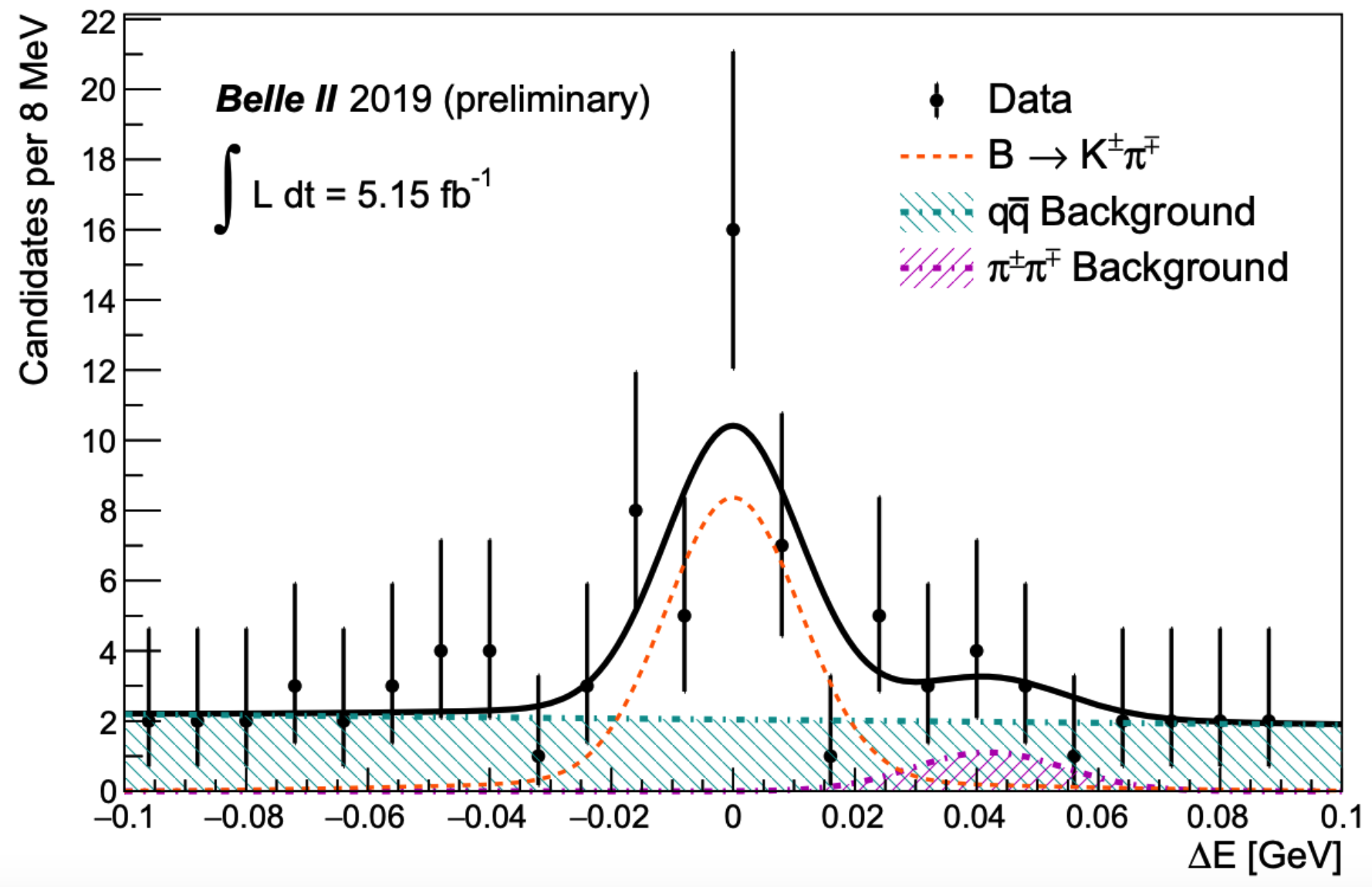} \\
\end{tabular}
\end{center}
 \caption{Distributions of $M_{bc}$ (left) and $\Delta E$ (right) for $B^{0}\to h^{+}{h}'^{-}$ ($h$, $h'$ = $\pi$ or $K$) candidates reconstructed in 5.15 fb$^{-1}$ of collision data with the projection of an unbinned maximum likelihood fit overlaid. The $M_{bc}$ distribution is obtained by restricting the data to the (-0.2, 0.2) GeV range of $\Delta E$; the $\Delta E$ distribution is obtained by restricting the data to the (5.275, 5.285) GeV/$c^{2}$ range of $M_{bc}$. A requirement on FBDT output suppresses continuum. Requirements on charged tracks PID suppress the combinatorial background.}
 \label{bkp}
\end{figure}

The sample of 5.15 fb$^{-1}$ taken by Belle II in a few months is already sufficient to reconstruct visible signals of charmless decays. However, a significant effort is needed to suppress continuum background. We target the $B^{0}\to K^{+}\pi^{-}$ decay, which has a quite large rate among charmless decays and is topologically straightforward. First, we select tracks satisfying simple track-quality and PID criteria. Then, a FBDT discriminator is applied to distinguish signal from continuum. The distributions of $M_{bc}$ and $\Delta E$ for $B^{0}\to h^{+}{h}'^{-}$ are shown in Fig. \ref{bkp}. The signal is dominated by approximately 25 $B^{0}\to K^{+}\pi^{-}$ decays, with a small indication of $B^{0}\to \pi^{+}\pi^{-}$ decays [5].

\section{Summary}

The first physics data corresponding to an integrated luminosity of 5.15 fb$^{-1}$ from the Belle II experiment are analyzed to validate the detector and software performance through reconstruction of various charmed and charmless $B$ decay modes. A total of approximately 4500 decays is reconstructed. This includes the observation of the Cabibbo-suppressed $B\to DK$ and the first reconstruction of the charmless $B^{0}\to K^{+}\pi^{-}$ signal in Belle II data. The results prove readiness for physics of the Belle II detector. \\ \\

\end{document}